\documentclass{article}
\usepackage{spconf,amsmath,graphicx}
\usepackage{subfigure}
\usepackage{graphicx}
\usepackage[table,xcdraw]{xcolor}
\usepackage{amsthm,amsmath,amssymb}
\usepackage{mathrsfs}
\usepackage{bbding}
\usepackage{pifont}
\usepackage{makecell}
\usepackage{url}

\title{AST-SED: an Effective Sound Event Detection Method Based on Audio Spectrogram Transformer}
\name{Kang Li$^1$, Yan Song$^1$, Li-Rong Dai$^1$, Ian McLoughlin$^{1,2}$, Xin Fang$^{3}$, Lin Liu$^{3}$}
\address{$^1$National Engineering Research Centre of Speech and Language Information Processing,\\
		University Of Science And Technology Of China, Hefei, China.\\
		$^2$ICT Cluster, Singapore Institute of Technology, Singapore.\\
		$^3$iFLYTEK Research, iFLYTEK Co. Ltd, Hefei, China.
}

\begin{document}
%\ninept
%
\maketitle
\begin{abstract}
	In this paper, we propose an effective sound event detection~(SED) method  based on the audio spectrogram transformer~(AST) model, pretrained on the large-scale AudioSet for audio tagging~(AT) task, termed AST-SED. 
	Pretrained AST models have recently shown promise on DCASE2022 challenge task4 where they help mitigate a lack of sufficient real annotated data.
	However, mainly due to differences between the AT and SED tasks, it is suboptimal to directly utilize outputs from a pretrained AST model.
	Hence the proposed AST-SED adopts an encoder-decoder architecture to enable effective and efficient fine-tuning without needing to redesign or retrain the AST model. %**IVM added "or retrain"
	Specifically, the Frequency-wise Transformer Encoder~(FTE) consists of transformers with self attention along the frequency axis to address multiple overlapped audio events issue in a single clip.
	The Local Gated Recurrent Units Decoder~(LGD) consists of nearest-neighbor interpolation~(NNI) and Bidirectional Gated Recurrent Units~(Bi-GRU) to compensate for temporal resolution loss in the pretrained AST model output.
	Experimental results on DCASE2022 task4 development set have demonstrated the superiority of the proposed AST-SED with FTE-LGD architecture.  
	Specifically, the Event-Based F1-score~(EB-F1) of 59.60\% and Polyphonic Sound detection Score scenario1~(PSDS1) score of 0.5140 significantly outperform CRNN and other pretrained AST-based systems. 
\end{abstract}
\section{Introduction}
\label{sec:intro}
    Sound Event Detection~(SED) aims at determining both the categories and timestamps of multiple overlapped events within a given audio clip.
    It has wide applications for real-world systems including smart home devices~\cite{southern2017sounding} and automatic surveillance~\cite{radhakrishnan2005audio}.
    Recently, DCASE challenges have set up a series of tasks\footnote{https://dcase.community/challenge2022/task-sound-event-detection-in-domestic-environments} to evaluate the progress of SED research. 
    Existing systems generally utilize semi-supervised learning methods, such as mean teacher~\cite{tarvainen2017mean,zheng2021effective,endo2022peer}, for weakly labeled SED tasks.
    Most of them exploit Convolutional Recurrent Neural Network~(CRNN) architectures~\cite{cakir2017convolutional,adavanne2017sound} and variants such as SK-CRNN~\cite{zheng2021improved} and FDY-CRNN~\cite{nam2022frequency} to perform frame-level feature extraction and context modeling for SED.
    %Miyazaki et. al~\cite{miyazaki2020conformer} introduced the Conformer architecture in 2020, enhancing Transformer attention-based blocks with convolutions, and utilizing mean-teacher semi-supervised learning to exploit strongly labeled and synthetic data to further improve SED performance. %**IVM I reworded this, please check it is correct because it is different from the original text
    In 2020, Miyazaki et. al~\cite{miyazaki2020conformer} introduced the Conformer, a convolution-augmented Transformer architecture , for modeling both local and global context information.
    With the help of strongly labeled and synthetic data, the SED performance can be further improved. 
    
\begin{figure}[t]
    	\centering
    	\subfigure[AST-GRU (baseline)]{
    		\begin{minipage}[t]{2.5in}
    			\centering
    			\includegraphics[width=\linewidth]{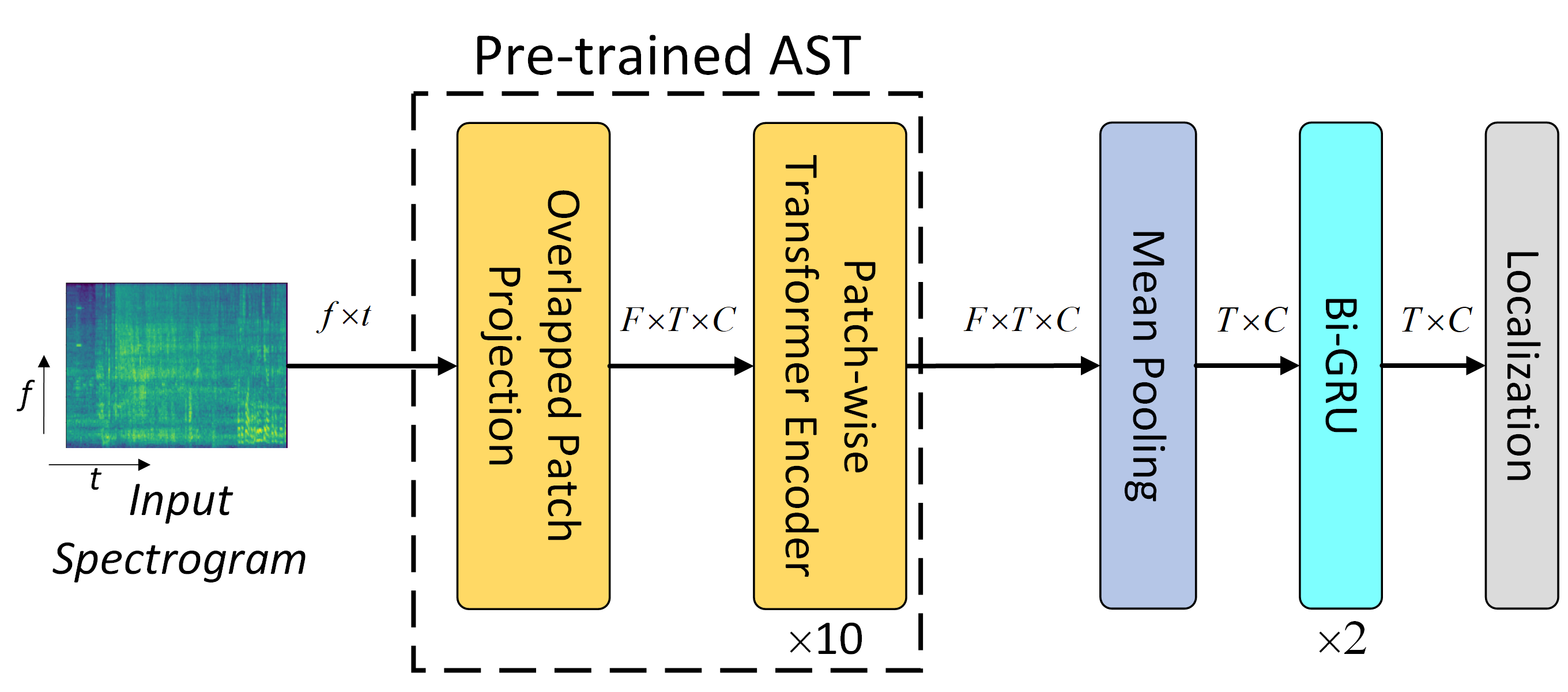}
    			\label{'fig:AST-GRU'}
    			%\caption{fig1}
    		\end{minipage}%
    	}%
    	
    	\subfigure[AST-SED (our proposed model)]{
    		\begin{minipage}[t]{3.3in}
    			\centering
    			\includegraphics[width=\linewidth]{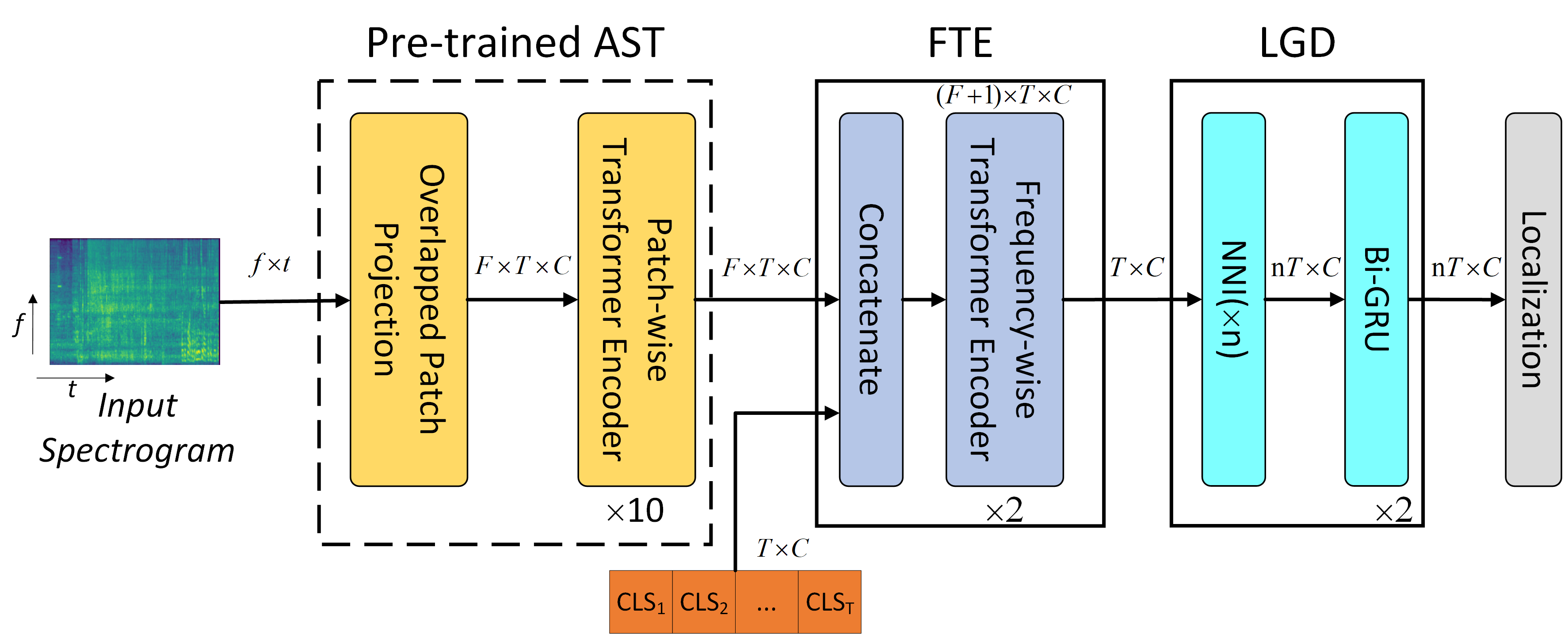}
    			\label{'fig:AST-SED'}
    			%\caption{fig2}
    		\end{minipage}%
    	}%
    	\caption{(a) AST-GRU: a frame-level sequence is obtained from the AST output by mean pooling along the frequency axis, feeding the Bi-GRU for context modeling.
    		(b) AST-SED: a frequency-wise transformer encoder~(FTE) derives a frame-level sequence, then Nearest Neighbour Interpolation~(NNI)  followed by Bi-GRUs are used as a local GRU decoder (LGD) to produce high temporal-resolution features.
    	}
    	\label{fig: models}
    \end{figure}

    In DCASE2022, several works were proposed to exploit external large-scale weakly-labeled AudioSet~\cite{gemmeke2017audio} data.
    For example, the forward-backward convolutional recurrent neural network (FB-CRNN) and Bi-directional CRNN (Bi-CRNN) of Ebbers and Haeb-Umbach~\cite{Ebbers2022}, are firstly pretrained on AudioSet, then fine-tuned for SED in a self-training manner.
    Xiao~\cite{Xiao2022} used RNNs to context model the output of the pretrained audio neural network~(PANN) and audio spectrogram transformer~(AST)~\cite{kong2020panns,gong2021ast}, both originally designed for audio tagging~(AT) tasks.
    Inspired by~\cite{Xiao2022}, we extend AST-GRU for SED using a pretrained AST model as shown in Fig.\ref{'fig:AST-GRU'}. Like a CRNN, the pretrained AST output is first aggregated (mean pooled) along the frequency axis to form a frame-level sequence, then fed to the Bi-GRU for context modeling.
    %%%%%%%%%%%%%%%%%%%%%%%%%%%%%%%%%%%%%%%
    %%%%%%%%%%%%%%%%%%%%%%%%%%%%%%%%%%%%%%%%	
    %Despite the promising performance achieved by using the pretrained AST model, it is still not optimal to %directly utilize the output of pretrained AST due to the difference between the AT and SED tasks.
    %%%%%%%%%%%%%%%%%%%%%%%%%%%%%%%%%%%%%%%%%%%%%%%%%%%%%%%%%%%%%%%%%%    
    \begin{figure}[t]
    	\centering
    	\subfigure[Mean pooling]{
    		\begin{minipage}[t]{0.5\linewidth}
    			\centering
    			\includegraphics[width=1in]{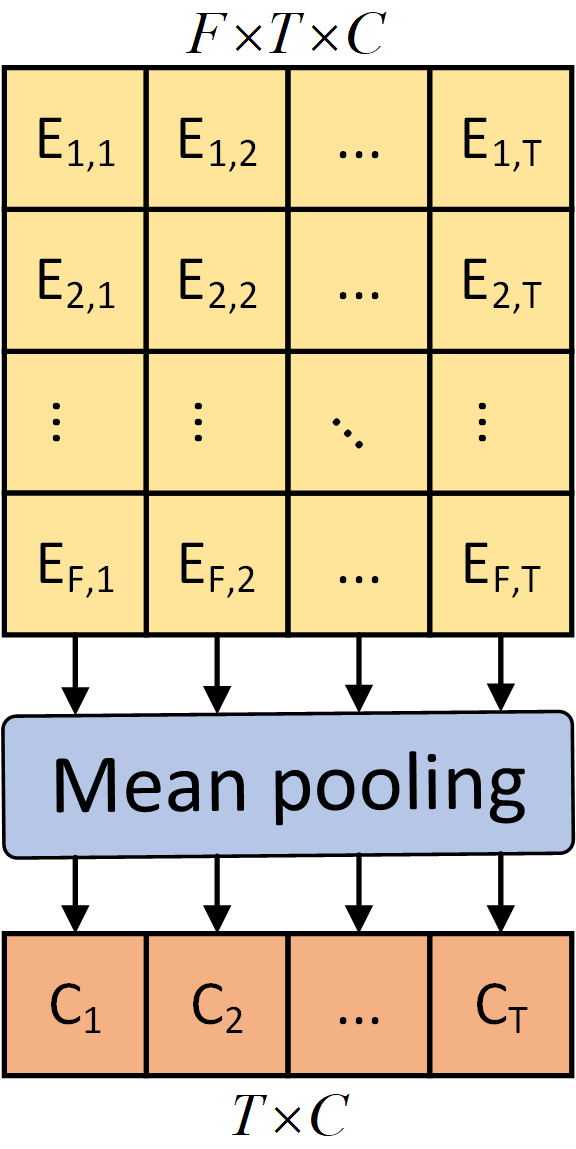}
    			\label{'fig:pwa'}
    			%\caption{fig1}
    		\end{minipage}%
    	}%
    	\subfigure[FTE with frequency-wise self attention]{
    		\begin{minipage}[t]{0.5\linewidth}
    			\centering
    			\includegraphics[width=1.6in]{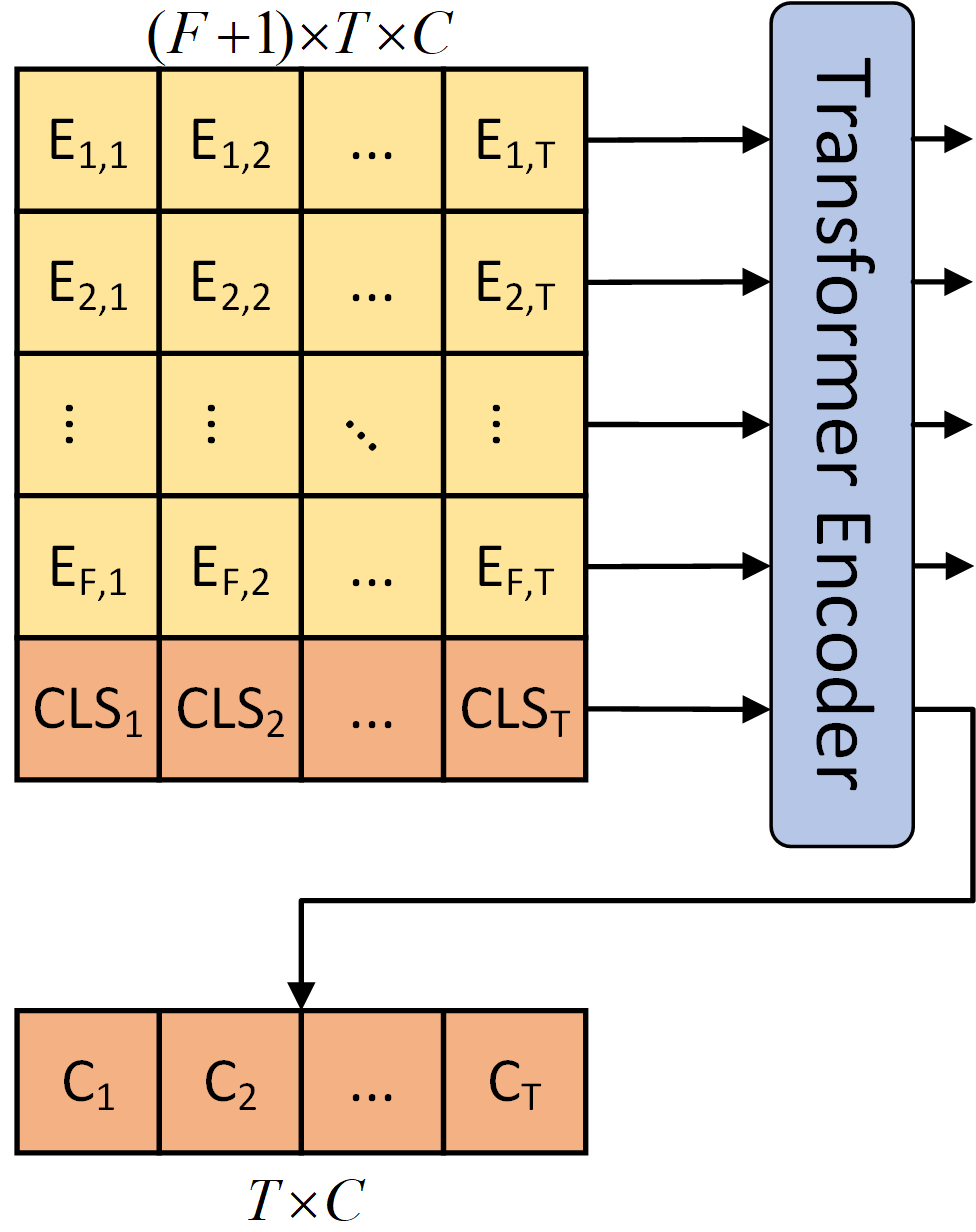}
    			\label{'fig:fwa'}
    			%\caption{fig2}
    		\end{minipage}%
    	}%
    	\caption{Comparison of Mean pooling in AST-GRU vs. FTE.
    		(a) mean pooling along frequency axis;
    		(b) FTE with self-attention along frequency axis.
    	}
    	\label{fig: self-attention}
    \end{figure}
%%%%%%%%%%%%%%%%%%%%%%%%%%%%%%%%%%%%%%%%%%%%%%%%%%%%%%%%%%%%%%%%%%    
    
    Despite the promising performance achieved by AST-GRU, it is still not optimal, as noted above, to directly utilize the output of a pretrained AST, due to differences between AT and SED tasks.
    First, there are multiple audio events with various frequency bands occurring in a single clip, so simple aggregation schemes (\textit{e.g.} mean pooling) may cause frequency information loss. 
    Second, the AST model is pretrained on AudioSet for AT tasks, which focuses on clip-wise representation, leading to potential temporal resolution loss.
    In this paper, we propose an effective AST-SED method with an encoder-decoder architecture based on an AST model pretrained on the large-scale AudioSet, as shown in Fig.\ref{'fig:AST-SED'}.
    Specifically, in the frequency-wise transformer encoder~(FTE),  frame-level class tokens are concatenated with AST outputs, followed by a series of transformer blocks. 
    Self-attention is performed along the frequency axis to obtain a frame-level representation suitable for multiple overlapped events.
    A local GRU decoder (LGD) first expands the frame sequence to match the temporal resolution of input spectrogram using nearest neighbour interpolation~(NNI), followed by a Bi-GRU~\cite{GRU} to perform context modeling for the SED task.
    As in previous systems~\cite{zheng2021effective}, mean teacher is used to learn from weakly labeled and unlabeled data.
    Extensive experiments have been conducted on the DCASE2022 challenge task4 development set to evaluate the proposed AST-SED with FTE-LGD architecture. 
    Specifically, a performance of 59.60\% EB-F1 and 0.5140 Polyphonic Sound detection Score scenario1~(PSDS1) is achieved, outperforming the 57.30\% and 0.5050 of the DCASE2022 winning system. 

\section{Method}
\label{sec:model}
	In this section, we briefly introduce the baseline AST-GRU system with pretrained AST, and then analyze its shortcomings.
	We then detail the AST-SED system with the proposed encoder-decoder (\textit{i.e.}, FTE-LGD) architecture for efficient fine-tuning without needing to redesign the AST model.
	
\subsection{Baseline system with the pretrained AST: AST-GRU}
	In~\cite{gong2021ast,gong2022ssast}, an AST architecture was proposed for AT tasks. The AST-GRU system is built upon the output of pretrained AST, as shown in Fig.\ref{'fig:AST-GRU'}.
	Input spectrograms are split into patches with size of 16$\times$16 and a stride of 10$\times$10. 
	These patches are linearly projected to 768-dimension patch-tokens.
	Following~\cite{dosovitskiy2020image,touvron2021training}, the patch-tokens plus positional embeddings, form the input sequences ${\textbf{PI}}\in\mathbb{R}^{F{\times}T{\times}C}$.
	The patch-wise transformer encoder~(PTE) with a multi-head self-attention (MHSA) is used to perform patch-level context modeling, as shown in Fig.\ref{'fig:AST-GRU'}. 
	
	It is known that the training of  transformer based models generally involves extensive computing resources, such as $O(n^2)$ computational complexity for self-attention spanning the length of patch sequences $n$,  and requires sufficient training data~\cite{touvron2021training}.
	This may limit the direct utilization of transformer based model for SED tasks, due to sequence length $n$, even when employing the large scale weakly labeled AudioSet.
%	This may limit  the directly utilization of transformer based model for SED task, even with large scale weakly labeled AudioSet. 
	In AST-GRU, mean pooling along the frequency axis is first applied to form frame sequences with much smaller $n$, followed by Bi-GRU for the SED task.
	A linear classifier with Sigmoid activation is used to produce frame-level SED predictions, and the softmax~\cite{wang2019comparison} is applied for the AT task.	
	Furthermore, the PTE part is initialized with the pretrained \textit{AST-base}, which has 10 transformer blocks.
	Following this, the AST-GRU is fine-tuned using Mean Teacher~\cite{tarvainen2017mean} for weakly-labeled and unlabeled data with the loss function defined as follows,
	\begin{equation} \label{1}
		L_{total}(t) = 0.5\cdot{L_{c}}+L_{f}+\alpha(t)\cdot(L_{MT,c}+L_{MT,f})
	\end{equation}
	where $L_{c}$, $L_{f}$ denote clip-level and frame-level classification loss, and $L_{MT,c}$, $L_{MT,f}$ denote clip-level and frame-level teacher-student consistency  loss respectively.
	Weight $\alpha(t)$ is tuned using a ramp-up scheme for each iteration $t$.
	Although AST-GRU can outperform existing CRNN architecture for SED, it is still not optimal mainly due to the difference of AT and SED tasks as aforementioned.
	In the next subsection, we will detail the proposed AST-SED system with FTE-LGD encoder-decoder architecture to address loss of frequency and temporal information in AST-GRU.
	
\subsection{AST-SED with proposed FTE-LGD architecture }
\subsubsection{FTE: Frequency-wise Transformer Encoder}
\begin{figure}[!htp]
	\centering
	\subfigure[Alarm\_bell\_ringing]{
		\begin{minipage}[t]{0.5\linewidth}
			\centering
			\includegraphics[width=1.5in]{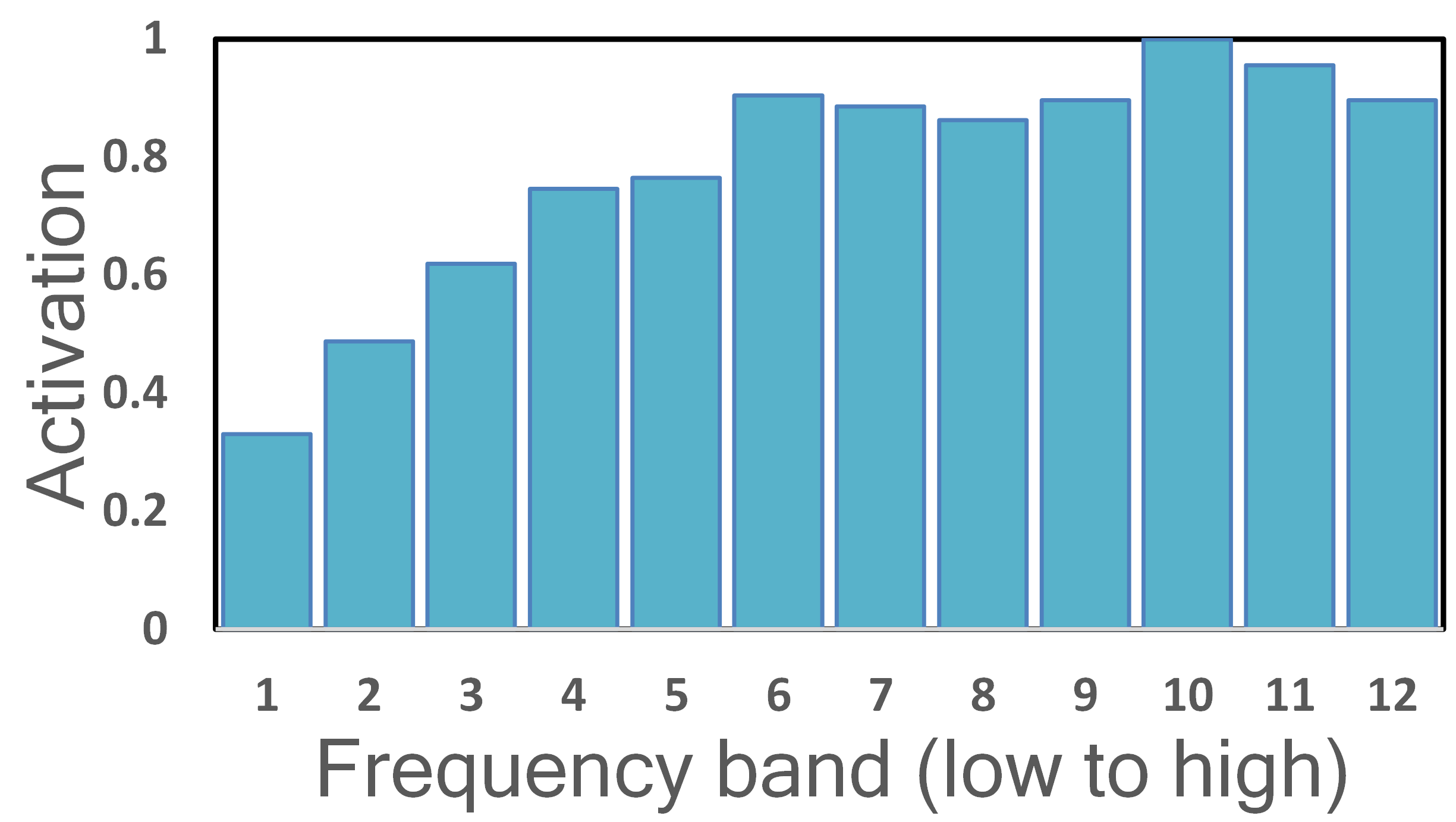}
			%\caption{fig1}
		\end{minipage}%
	}%
	%\quad
	%\centering
	\subfigure[Dog]{
		\begin{minipage}[t]{0.5\linewidth}
			\centering
			\includegraphics[width=1.5in]{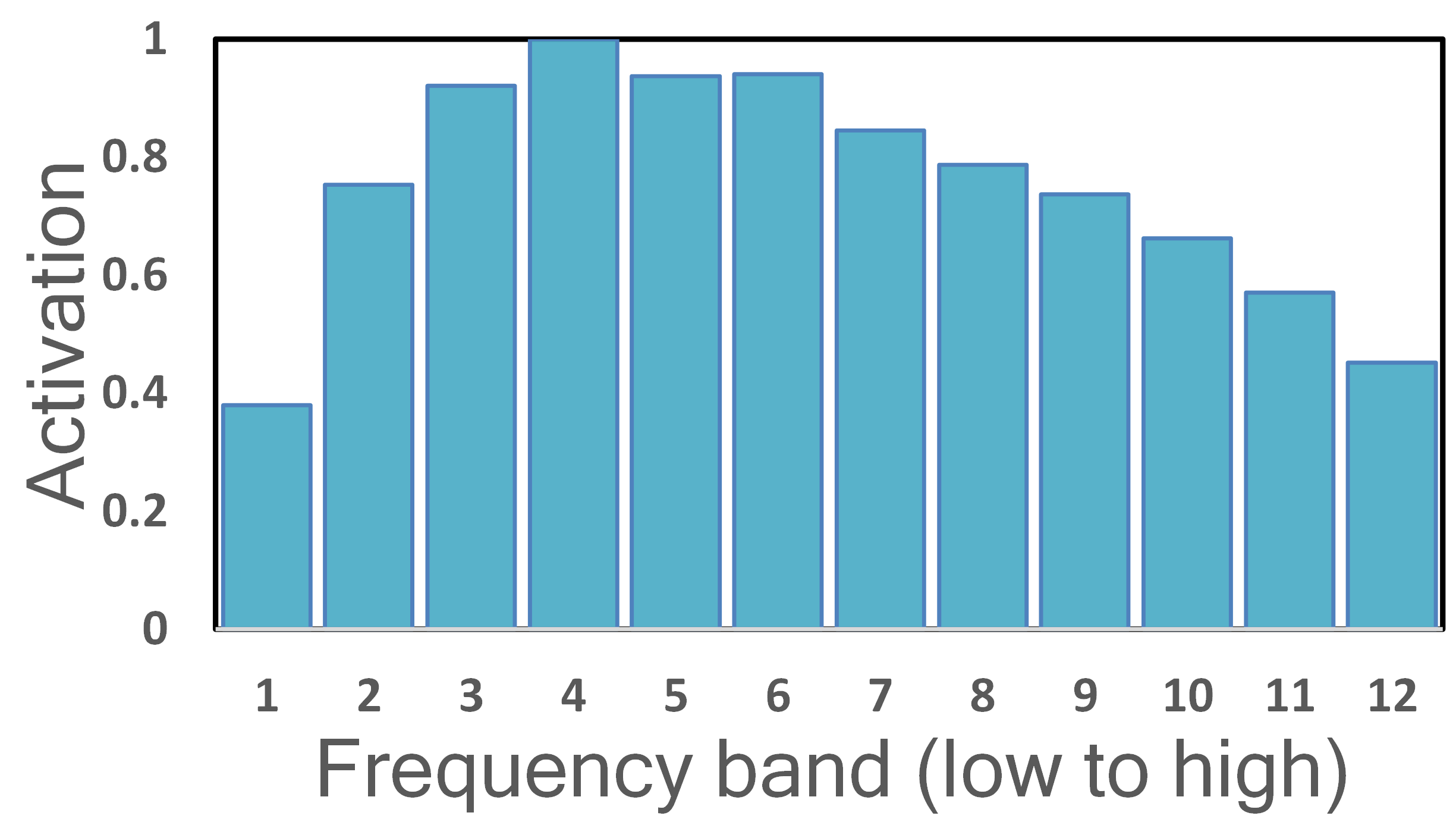}
			%\caption{fig2}
		\end{minipage}%
	}%
	%\quad
	%\centering
	
	\subfigure[Blender]{
		\begin{minipage}[t]{0.5\linewidth}
			\centering
			\includegraphics[width=1.5in]{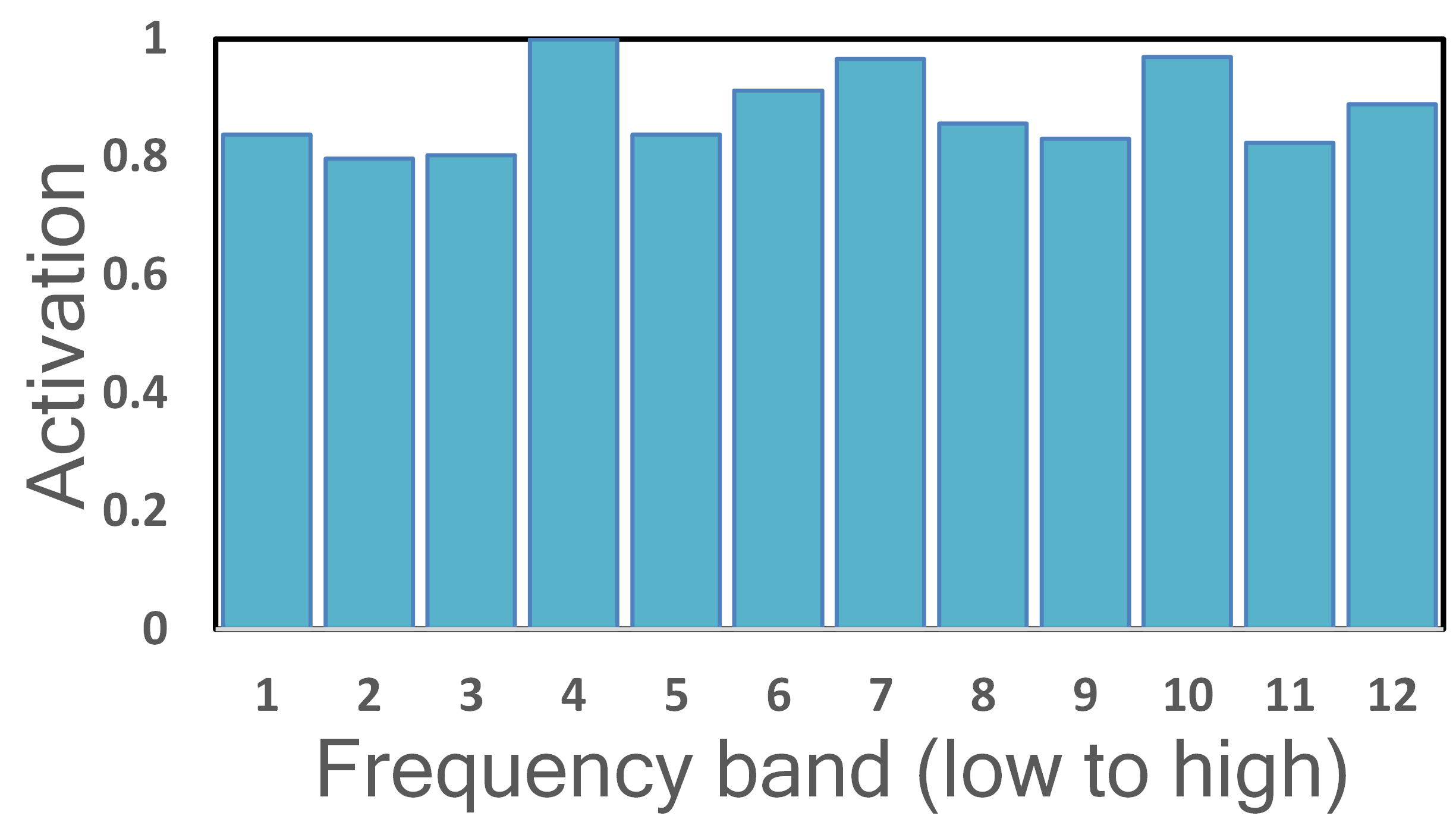}
			%\caption{fig2}
		\end{minipage}
	}%
	%\quad
	%\centering
	\subfigure[Speech]{
		\begin{minipage}[t]{0.5\linewidth}
			\centering
			\includegraphics[width=1.5in]{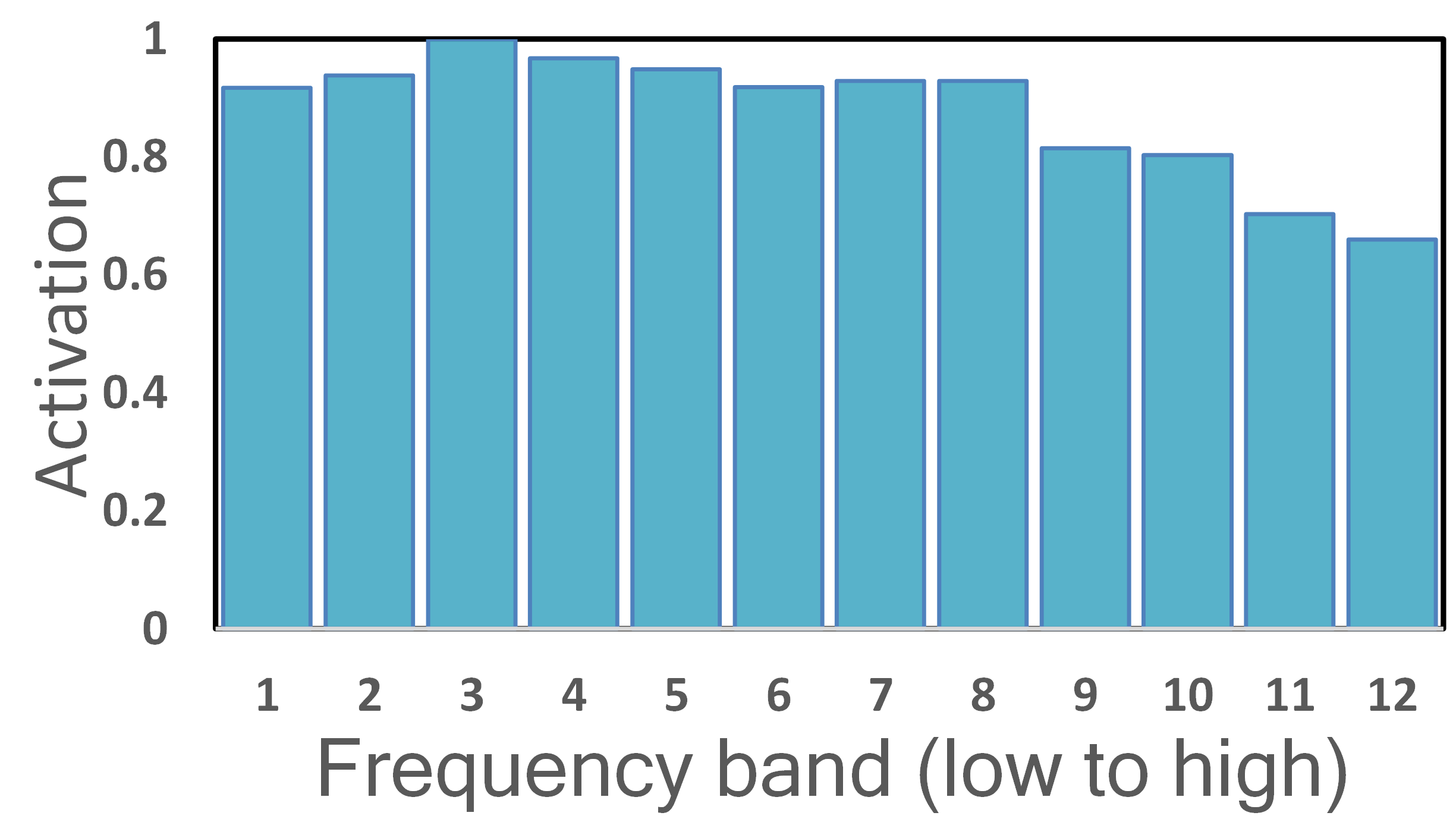}
			%\caption{fig2}
		\end{minipage}
	}%
	%\quad
	\centering
	\caption{Normalized activation distribution along frequency dimension.}
	\label{fig: dis}
\end{figure}

	Let us first evaluate the frequency-wise activation distribution of each event to illustrate the motivation for FTE.
	This is done by replacing the mean pooling in AST-GRU with a single frequency band, to evaluate the detection performance of each event category.
	Class-wise EB-F1 using different frequency bands is calculated and normalized to $[0,1]$ to reflect the frequency-wise activation distribution.
	Fig.\ref{fig: dis} presents histograms to explore the frequency-wise distribution of four kinds of events:
	(1) \textit{Alarm\_bell\_ringing} events that mainly activate on a high frequency band, 
	(2) \textit{Speech} events that mainly activate on low frequency band, 
	(3) \textit{Dog}, \textit{Dishes} and \textit{Cat} events \textit{etc.,} that mainly activate on middle frequencies, and
	(4) \textit{Electronic\_shaver}, \textit{Blender}, \textit{Vacuum\_cleaner} and \textit{Frying} events \textit{etc.} that activate on all frequency bands.
	We observe differing distributions with respect to frequency, suggesting that simple aggregation schemes like mean pooling may inevitably cause loss of frequency information that is potentially discriminative for SED. 
	
	To address this, an FTE block is proposed to introduce frequency-wise MHSA~(fMHSA) in transformer.
	Specifically, given the PTE output ${\textbf{PO}}\in\mathbb{R}^{F{\times}T{\times}C}$, a learnable \textbf{CLS} token sequence $\textbf{CLS}\in\mathbb{R}^{T{\times}C}$ is concatenated with $\textbf{PO}$ along the frequency axis to form the FTE input $\textbf{FI}\in\mathbb{R}^{{(F+1)\times}T{\times}C}$.
	Following~\cite{dosovitskiy2020image,touvron2021training}, an FTE block mainly consists of  multi-layer perceptions~(MLP) with ~fMHSA, followed by layer normalization~(LN) as follows,
	\begin{eqnarray}
		\hat{\textbf{FI}} = LN(fMHSA(\textbf{FI})+\textbf{FI})		\\
		{\textbf{FO}} = LN(MLP(\hat{\textbf{FI}})+\hat{\textbf{FI}})
	\end{eqnarray}
	It should be noted that in fMHSA, each token $E_{t,f}$, only interacts with other tokens $E_{t,\cdot}$, as illustrated in Fig.\ref{'fig:fwa'}.
	The $\textbf{CLS}$ tokens from FTE output $\textbf{FO}\in\mathbb{R}^{(F+1){\times}T{\times}C}$ forms the final frame sequence $\textbf{C}\in\mathbb{R}^{T{\times}C}$.
	The FTE comprises of 2 transformer blocks with 4-head fMHSA, and the intermediate embedding dimension is set to 768.
	
%%%%%%%%%%%%%%%%%%%%%%%%%%%%%%%%%%%%%%%%%%%%%%%%%%%%%%%%%%%%%%
%%%%%%%%%%%%%%%%%%%%%%%%%%%%%%%%%%%%%%%%%%%%%%%%%%%%%%%%%%%%%%

\subsubsection{LGD: Local GRU Decoder}
	Since the AST output has low temporal resolution, directly feeding it to the Bi-GRU for context modeling may achieve limited SED performance.
	%LGD therefore contains an NNI block before the Bi-GRU as a local decoder, applied on the FTE output sequence $\textbf{C}\in\mathbb{R}^{T{\times}C}$, as shown in Fig.~\ref{'fig:AST-SED'}.
	LGD therefore contains an NNI block and the Bi-GRU to form a local decoder, applied on the FTE output sequence $\textbf{C}\in\mathbb{R}^{T{\times}C}$, as shown in Fig.\ref{'fig:AST-SED'}.
    If the NNI up-sampling ratio (UR) is denoted as $n$, a simple implementation of LGD is as follows:
	\begin{eqnarray}
		\hat{\textbf{C}} = NNI(\textbf{C},n) 	\\
		\textbf{O} = BiGRU(\hat{\textbf{C}}) 	
	\end{eqnarray}
	The output $\textbf{O}\in\mathbb{R}^{nT{\times}C}$ may be suitable for a fine-grained SED task with high temporal resolution.
	The decoding ability of GRUs~\cite{GRU} has largely been ignored in CRNNs because the CNN part is designed to produce high-temporal-resolution output, with no need to decode. 
	However, in AST-based SED, it is inefficient to use the high temporal resolution transformer for pretraining or fine tuning due to the extensive computing resource, including labeled data and computation costs~\cite{touvron2021training}.
	This is also the motivation to choose a Bi-GRU for SED, instead of Transformer,  in the Conformer~\cite{miyazaki2020conformer}.

%	(1) transformer need more training data to unlock its power, therefore AST is firstly initialed with pretrained DEiT, then pretrained on AudioSet, it must follow same low-resolution design as DEiT \cite{touvron2021training} and is suitable for AT task.
%	(2) the computation complexity of AST is $O(n^2)$, high resolution means big $n$ which results high computation cost.
%	Introducing a decoder works on AST output to recover resolution is a good solution for SED task.
\section{Experiments and Results}
\label{sec:exp}
\subsection{Datasets and Experiment Setup}
	For evaluation, we employ the DCASE2022 task4 development set (DESED)~\cite{turpault2019sound}. 
	The training dataset contains 1578 weak clips, 14412 unlabeled clips, 3470 real-strong clips and 10000 synthetic-strong clips.
	The validation dataset consists of 1168 clips.
     PSDS1~\cite{PSDS} and EB-F1 scores are used to evaluate  fine-grained SED performance.

	A 16kHz audio input waveform is first converted into 128-dimensional log Mel spectrogram features with a window size of 25ms and frame shift of 10ms, and then normalized to mean 0 and standard deviation 0.5.
	Data augmentations such as MixUp~\cite{zhang2018mixup}, time-mask, time-shift and FilterAugment~\cite{nam2022filteraugment} are used.
	Median filters with fixed time length are used for post-processing, and all event types share a threshold of 0.5 to obtain hard predictions for calculating EB-F1.
	The data augmentation and Median filter parameters are identical to~\cite{nam2022frequency}. 
	Both AST-GRU and AST-SED are trained over 10 epochs using the AdamW optimizer~\cite{AdamW}, and a ratio of 1:1:2:2 for real-strong, real-synthetic, weak and unlabeled data is used for each batch. 
	%Learning rates~(lr) are set to 5e-6, 1e-4 for pre-trained PTE and the remaining parts respectively. 
	Learning rates~(lr) are set to 5e-6, 1e-4 for pre-trained AST and the remaining parts respectively. 
	During training, the lr is constant for the first 5 epochs, then reduced exponentially.
	Mean teacher SSL is exploited for fine-tuning the AST-GRU and AST-SED using DCASE2022 DESED.

\subsection{AST-GRU and AST-SED performance compared}
	As shown in Table~\ref{tab:1}, AST-GRU achieves an EB-F1 of 55.20\% and PSDS1 of 0.4537 on DCASE 2022 DESED, significantly  outperforming the   CRNN baseline scores of 50.5\% and 0.4006 respectively.
	This demonstrates that an AST pretrained on AudioSet is a strong feature extractor under the same experimental settings. 
	The AST-SED with FTE-LGD encoder-decoder architecture can achieve even better results. 
	Specifically, the highest EB-F1 of 59.60\% and PSDS1 of 0.5140 can be obtained, outperforming recently reported state-of-the-art results. 
	It is worth noting that the previous winning system~\cite{Ebbers2022} was trained in a self-training manner.
	
%%%%%%%%%%%%%%%%%%%%%%%%%%%%%%%%%%%%%%
\begin{table}[t]
	\centering
	\caption{Comparison of model performance on DCASE 2022 DESED. {$\dag$} denotes results from our implementation using the codebase from~\cite{nam2022frequency}. }
	\setlength{\tabcolsep}{1mm}
	\begin{tabular}{lcc}
		\hline
		\textbf{Model}                    & \textbf{EB-F1,\%}  & \textbf{PSDS1}  \\ \hline
		CRNN{$^\dag$}                 & 50.50 & 0.4006 \\
		FDY-CRNN{$^\dag$}  \cite{nam2022frequency}             & 51.56 & 0.4256 \\
		SK-CRNN{$^\dag$} \cite{zheng2021improved}               & 52.77 & 0.4004 \\
		Ensembled PANNs-RNN \cite{Xiao2022}             &N/A&0.4450\\
		Ensembled AST-RNN \cite{Xiao2022}                 &N/A&0.4590\\
		BiCRNN (Winner) \cite{Ebbers2022}           & 57.30 & 0.5050 \\
		AST-GRU{$^\dag$} (ours)                   & \textbf{55.20} & \textbf{0.4537} \\ 
		AST-SED{$^\dag$} (ours)   & \textbf{59.60} & \textbf{0.5140} \\ \hline	
	\end{tabular}
	\label{tab:1}
\end{table}
%%%%%%%%%%%%%%%%%%%%%%%%%%%%%%%%%%%%%%
\begin{table}[]
	\centering
	\caption{Performance of the proposed method on DCASE 2022 DESED. UR denotes up-sampling ratio.}
	\setlength{\tabcolsep}{1mm}
	\begin{tabular}{lcc}
		\hline
		\textbf{Model}             	& \textbf{EB-F1,\%}  & \textbf{PSDS1}  \\ \hline
%		CNN-GRU (CRNN) 				& 50.50 & 0.4006 \\
		AST-GRU                		& 55.20 & 0.4537 \\
		AST-GRU+FTE  				& 57.86 & 0.4742 \\  
		AST-GRU+LGD   				& 57.50 & 0.4933 \\ 
		AST-GRU+FTE+LGD (AST-SED)   & \textbf{59.60} & \textbf{0.5140} \\\hline
		AST-SED (UR=1)              & 57.86 & 0.4742 \\
		AST-SED (UR=2)             & 59.12 & 0.5016 \\
		AST-SED (UR=5)          	& \textbf{60.00} & 0.5110 \\
		AST-SED (UR=10) 		& 59.60 & \textbf{0.5140} \\  \hline
	\end{tabular}
	\label{tab:2}
\end{table}
%%%%%%%%%%%%%%%%%%%%%%%%%%%%%%%%%%%%%%
\begin{table}[!tb]
	\centering
	\caption{Relative improvement (RI) for individual events  with LGD;
		S: Speech,
		D: Dog,
		R: Running\_water,
		E: Electric\_shaver\_toothbrush,
		A: Alarm\_bell\_ringing,
		C: Cat,
		F: Frying,
		B: Blender,
		V: Vacuum\_cleaner,
		Di: Dishes.
		%Short and long clips are also compared.
		Short and long event sets are also compared.
	}
%%%%%%%%%%%%%%%%%%%%%%%%%%%%%%%%%%
	\setlength{\tabcolsep}{1.8mm}
	\begin{tabular}{lcccc}
		\hline
		\textbf{Event}       & \textbf{\makecell[c]{Mean\\duration, s}} & \textbf{\makecell[c]{EB-F1, \%\\(w/o)}} & \textbf{\makecell[c]{EB-F1, \%\\(w)}} & \textbf{RI, \%} \\ \hline
		S                     & 1.50    & 60.27          & 66.07        & \textbf{9.62}                     \\
		D                       & 1.37    & 42.30          & 46.37        & \textbf{9.61}                     \\
		R                       & 5.15    & 56.73          & 56.80        & 0.12                     \\
		E                       & 7.90    & 71.27          & 74.53        & 4.58                     \\
		A                       & 1.96    & 54.77          & 55.23        & 0.85                     \\
		C                       & 1.39    & 52.37          & 54.00        & 3.12                     \\
		F                       & 8.23    & 67.83          & 71.73        & 5.75                     \\
		B                     & 5.15    & 61.70          & 64.60        & 4.70                     \\
		V                       & 8.45    & 75.30          & 69.80        & -7.30                    \\
		Di                     & 0.63    & 36.07          & 36.80        & 2.03                     \\ \hline
		Short & 0-2       & 49.15          & 51.69        & \textbf{5.17}                     \\
		Long  & 2-10       & 66.57          & 67.49        & 1.39                     \\
		All                    & -       & 57.86          & 59.59        & 3.00                     \\ \hline
	\end{tabular}
	\label{tab:3}
\end{table}
%%%%%%%%%%%%%%%%%%%%%%%%%%%%%%%%%%

\subsection{AST-SED with FTE-LGD ablation study}
	We conduct an ablation study consisting of a series of experiments to evaluate the effectiveness of different architecture configurations and parameter choices. 
	Table~\ref{tab:2} presents configurations with FTE or LGD which may outperform the baseline AST-GRU. 
	The AST-GRU/SED with FTE-LGD encoder-decoder architecture achieves the best performance, with an absolute improvement over AST-GRU of 4.40\% in EB-F1 and 0.06 in PSDS1.
	We also consider LGD performance with various up-sample ratios in Table~\ref{tab:2}, revealing that ratio in of 5 and 10 perform best.

	Table~\ref{tab:3} separates performance on different audio event types. For short duration events, such as \textit{Speech} and \textit{Dog}, AST-SED obtains a relatively significant improvement of 9.62\% and 9.61\% respectively.
	Long duration events, such as \textit{Running\_water} and \textit{Vacuum\_cleaner}, obtain more limited gains from AST-SED.
	This may attribute to the relatively simple NNI operation in LGD. 
	In the future, we aim to explore more powerful decoders to better benefit longer event types. 
		
\section{Conclusion}
\label{sec:cl}
	This paper has proposed an effective AST-SED method based on a pretrained AST model.
	Specifically, an encoder-decoder architecture was designed, without needing to revise the AST model, for efficient training.
	The encoder FTE, consisting of transformer blocks with frequency-wise self attention, is used for effective frame-wise feature extraction.
	Meanwhile the decoder LGD, consisting of bi-GRU blocks with an NNI operation, is applied top achieved fine-grained capabilities for SED.
	Experimental results on DCASE2022  task4 demonstrate the capabilities of the proposed AST-SED with FTE-LGD architecture to outperform state-of-the-art methods.
	
\bibliographystyle{IEEEbib}
\bibliography{strings,refs}

\begin{thebibliography}{10}
\setlength{\itemsep}{0pt}
\bibitem{southern2017sounding}
A.~Southern, F.~Stevens, and D.~Murphy,
\newblock ``Sounding out smart cities: Auralization and soundscape monitoring
  for environmental sound design,''
\newblock {\em J. Acoustical Soc. America}, vol. 141, no. 5, pp. 3880--3880,
  2017.

\bibitem{radhakrishnan2005audio}
R.~Radhakrishnan, A.~Divakaran, and A.~Smaragdis,
\newblock ``Audio analysis for surveillance applications,''
\newblock in {\em IEEE WASPAA 2005.}, 2005, pp. 158--161.

\bibitem{tarvainen2017mean}
A.~Tarvainen and H.~Valpola,
\newblock ``Mean teachers are better role models: Weight-averaged consistency
  targets improve semi-supervised deep learning results,''
\newblock {\em Advances in neural information processing systems}, vol. 30,
  2017.

\bibitem{zheng2021effective}
X.~Zheng, Y.~Song, L.-R. Dai, I.~McLoughlin, and L.~Liu,
\newblock ``An effective mutual mean teaching based domain adaptation method
  for sound event detection.,''
\newblock in {\em Interspeech}, 2021, pp. 556--560.

\bibitem{endo2022peer}
H.~Endo and H.~Nishizaki,
\newblock ``Peer collaborative learning for polyphonic sound event detection,''
\newblock in {\em IEEE ICASSP}, 2022, pp. 826--830.

\bibitem{cakir2017convolutional}
E.~Cak{\i}r, G.~Parascandolo, T.~Heittola, H.~Huttunen, and T.~Virtanen,
\newblock ``Convolutional recurrent neural networks for polyphonic sound event
  detection,''
\newblock {\em IEEE Trans. Audio, Speech, and Language Proc.}, vol. 25, no. 6,
  pp. 1291--1303, 2017.

\bibitem{adavanne2017sound}
S.~Adavanne, P.~Pertil{\"a}, and T.~Virtanen,
\newblock ``Sound event detection using spatial features and convolutional
  recurrent neural network,''
\newblock in {\em IEEE ICASSP}, 2017, pp. 771--775.

\bibitem{zheng2021improved}
X.~Zheng, Y.~Song, I.~McLoughlin, L.~Liu, and L.-R. Dai,
\newblock ``An improved mean teacher based method for large scale weakly
  labeled semi-supervised sound event detection,''
\newblock in {\em IEEE ICASSP}, 2021, pp. 356--360.

\bibitem{nam2022frequency}
H.~Nam, S.-H. Kim, B.-Y. Ko, and Y.-H. Park,
\newblock ``Frequency dynamic convolution: Frequency-adaptive pattern
  recognition for sound event detection,''
\newblock {\em arXiv preprint arXiv:2203.15296}, 2022.

\bibitem{miyazaki2020conformer}
K.~Miyazaki, T.~Komatsu, T.~Hayashi, S.~Watanabe, T.~Toda, and K.~Takeda,
\newblock ``Convolution augmented transformer for semi-supervised sound event
  detection,''
\newblock in {\em Proc. Workshop Detection Classification Acoust. Scenes Events
  (DCASE)}, 2020, pp. 100--104.

\bibitem{gemmeke2017audio}
J.~F. Gemmeke, D.~P. Ellis, D.~Freedman, A.~Jansen, W.~Lawrence, R.~C. Moore,
  M.~Plakal, and M.~Ritter,
\newblock ``Audio set: An ontology and human-labeled dataset for audio
  events,''
\newblock in {\em IEEE ICASSP}, 2017, pp. 776--780.

\bibitem{Ebbers2022}
J.~Ebbers and R.~Haeb-Umbach,
\newblock ``Pre-training and self-training for sound event detection in
  domestic environments,''
\newblock Tech. {R}ep., DCASE, June 2022.

\bibitem{Xiao2022}
S.~Xiao,
\newblock ``Pretrained models in sound event detection for dcase 2022 challenge
  task4,''
\newblock Tech. {R}ep., DCASE, June 2022.

\bibitem{kong2020panns}
Q.~Kong, Y.~Cao, T.~Iqbal, Y.~Wang, W.~Wang, and M.~D. Plumbley,
\newblock ``Panns: Large-scale pretrained audio neural networks for audio
  pattern recognition,''
\newblock {\em IEEE Trans. Audio, Speech, and Language Proc.}, vol. 28, pp.
  2880--2894, 2020.

\bibitem{gong2021ast}
Y.~Gong, Y.-A. Chung, and J.~Glass,
\newblock ``{AST: Audio Spectrogram Transformer},''
\newblock in {\em Proc. Interspeech 2021}, 2021, pp. 571--575.

\bibitem{GRU}
K.~Cho, B.~Van~Merri{\"e}nboer, C.~Gulcehre, D.~Bahdanau, F.~Bougares,
  H.~Schwenk, and Y.~Bengio,
\newblock ``Learning phrase representations using rnn encoder-decoder for
  statistical machine translation,''
\newblock {\em arXiv preprint arXiv:1406.1078}, 2014.

\bibitem{gong2022ssast}
Y.~Gong, C.-I. Lai, Y.-A. Chung, and J.~Glass,
\newblock ``Ssast: Self-supervised audio spectrogram transformer,''
\newblock in {\em Proceedings of the AAAI Conference on Artificial
  Intelligence}, 2022, vol.~36, pp. 10699--10709.

\bibitem{dosovitskiy2020image}
A.~Dosovitskiy, L.~Beyer, A.~Kolesnikov, D.~Weissenborn, X.~Zhai,
  T.~Unterthiner, M.~Dehghani, M.~Minderer, G.~Heigold, S.~Gelly, J.~Uszkoreit,
  and N.~Houlsby,
\newblock ``An image is worth 16x16 words: Transformers for image recognition
  at scale,''
\newblock {\em ICLR}, 2021.

\bibitem{touvron2021training}
H.~Touvron, M.~Cord, M.~Douze, F.~Massa, A.~Sablayrolles, and H.~J{\'e}gou,
\newblock ``Training data-efficient image transformers \& distillation through
  attention,''
\newblock in {\em International Conference on Machine Learning}. PMLR, 2021,
  pp. 10347--10357.

\bibitem{wang2019comparison}
Y.~Wang, J.~Li, and F.~Metze,
\newblock ``A comparison of five multiple instance learning pooling functions
  for sound event detection with weak labeling,''
\newblock in {\em IEEE ICASSP}, 2019, pp. 31--35.

\bibitem{turpault2019sound}
N.~Turpault, R.~Serizel, J.~Salamon, and A.~P. Shah,
\newblock ``Sound event detection in domestic environments with weakly labeled
  data and soundscape synthesis,''
\newblock 2019.

\bibitem{PSDS}
{\c{C}}.~Bilen, G.~Ferroni, F.~Tuveri, J.~Azcarreta, and S.~Krstulovi{\'c},
\newblock ``A framework for the robust evaluation of sound event detection,''
\newblock in {\em IEEE ICASSP}, 2020, pp. 61--65.

\bibitem{zhang2018mixup}
Y.~N. D. D. L.-P. Hongyi~Zhang, Moustapha~Cisse,
\newblock ``mixup: Beyond empirical risk minimization,''
\newblock {\em International Conference on Learning Representations}, 2018.

\bibitem{nam2022filteraugment}
H.~Nam, S.-H. Kim, and Y.-H. Park,
\newblock ``Filteraugment: An acoustic environmental data augmentation
  method,''
\newblock in {\em IEEE ICASSP}, 2022, pp. 4308--4312.

\bibitem{AdamW}
I.~Loshchilov and F.~Hutter,
\newblock ``Decoupled weight decay regularization,''
\newblock {\em arXiv preprint arXiv:1711.05101}, 2017.

\end{thebibliography}
\end{document}